\documentstyle[12pt,aaspp4,epsf,flushrt]{article}

\newcommand{\be}{\begin{equation}}
\newcommand{\ee}{\end{equation}}
\newcommand{\half}{{\textstyle{1\over2}}}

\def\spose#1{\hbox to 0pt{#1\hss}}
\def\lta{\mathrel{\spose{\lower 3pt\hbox{$\mathchar"218$}}
     \raise 2.0pt\hbox{$\mathchar"13C$}}}
\def\gta{\mathrel{\spose{\lower 3pt\hbox{$\mathchar"218$}}
     \raise 2.0pt\hbox{$\mathchar"13E$}}}
 
\def\Re{\mathop{\it Re}\nolimits}
\def\Im{\mathop{\it Im}\nolimits}

\def\pdrv#1#2{{\partial #1 \over \partial #2}}
\def\drv#1#2{{d #1 \over d #2}}
\def\tdrv2#1#2{{d^2 #1\over d{#2}^2}}

\def\LBEreal{\pdrv{f_1}{t}+\vec v\cdot\nabla f_1
        -\nabla\Phi_0\cdot\pdrv{f_1}{\vec v}=
        \nabla\Phi_1\cdot\pdrv{f_0}{\vec v}}
\def\Phiav {\langle\Phi_1\rangle}
\def\llp {\ell(\ell+1)}

\def\bfr{{\bf r}}
\def\bfv{{\bf v}}
\def\bfw{{\bf w}}
\def\bfk{{\bf k}}
\def\bfOmega{{\bf \Omega}}
\def\lm{{\ell m}}
\def\tPhi{\widetilde \Phi}
\def\bfI{{\bf I}}
  \font\gkvec=cmmib10  
\def\bfomega{\hbox{{\gkvec\char33}}}
\def\bnabla{\mbox{\boldmath $\nabla$}}
\def\omit#1{}
\def\mathnew{\mathsurround=0pt}
\def\simov#1#2{\lower .5pt\vbox{\baselineskip0pt \lineskip-.5pt
        \ialign{$\mathnew#1\hfil##\hfil$\crcr#2\crcr\sim\crcr}}}

\def\simless{\mathrel{\mathpalette\simov <}}

\def\LBEreal{\drv{f}{t}\equiv\pdrv{f}{t}+\bfv\cdot\pdrv{f}{\bfr}
	-\bnabla\Phi_0\cdot\pdrv{f}{\bfv}=
	\bnabla\Phi_1\cdot\pdrv{F}{\bfv}}

\begin{document}

\title{Linear response of galactic halos to adiabatic 
	gravitational perturbations}

\author{Chigurupati Murali$^1$ and Scott Tremaine$^{1,2}$}

\affil{$^1$Canadian Institute for Theoretical Astrophysics, McLennan Labs, 
University of Toronto,\\ 60~St.\ George St., Toronto M5S 3H8, Canada}

\medskip

\affil{$^2$Canadian Institute for Advanced Research, Program in Cosmology and
Gravity}

\begin{abstract}
\noindent
We determine the response of a self-similar isothermal stellar system to small
adiabatic gravitational perturbations.  For odd spherical harmonics, the
response is identical to the response of the analogous isothermal fluid
system. For even spherical harmonics, the response can be regarded as an
infinite series of wavetrains in $\log r$, implying alternating compression
and rarefaction in equal logarithmic radius intervals. Partly because of the
oscillatory nature of the solutions, tidal fields from external sources are
not strongly amplified by an intervening isothermal stellar system, except at
radii $\lta 10^{-3.5}$ times the satellite radius; at some radii the stellar
system can even screen the external tidal field in a manner analogous to Debye
screening. As Weinberg has pointed out, individual resonances in a stellar
system can strongly amplify external tidal fields over a limited radial range,
but we cannot address this possibility because we examine only adiabatic
perturbations.  We also discuss the application of our method to the halo
response caused by the slow growth of an embedded thin disk.
\end{abstract}


\section{Introduction}
\label{sec:intro}

\noindent	
In models of galaxy formation based on hierarchical clustering, the
gravitational field becomes less and less smooth at larger and larger
distances from the galaxy center. The halos of isolated galaxies contain
satellite and companion galaxies and merging dark matter substructure, and
galaxies in groups and clusters are subject to slowly varying tidal fields
from distant group members as well as rapidly changing forces from close
encounters. Despite this noisy environment, the visible inner regions of most
galaxies are relatively smooth and regular.

One expects that gravitational noise from the outer halo and beyond
will not severely disturb the inner galaxy because the dominant
(quadrupole) tidal potential at radius $r$ from a source of mass $m_p$
at radius $r_p\gg r$ varies as $Gm_pr^2/r_p^3$; in other words the
ratio of the tidal force to the force from the body of the galaxy
itself is of order $(m_p/m)(r/r_p)^3$ where $m$ is the mass of the
galaxy within $r$. The cubic dependence of the force ratio on the
radius ratio implies that tidal forces from distant satellites are
generally unable to strongly perturb the inner galaxy ($r\ll r_p$).

However, this assessment neglects an important effect: in a realistic galaxy
model the tidal forces do not propagate through a vacuum---rather, they
propagate through the density field of the dark-matter halo.  The halo
provides an intervening medium that modifies the external tidal field in a
manner analogous to the polarization of a dielectric medium in electrostatics
(Jackson 1975). Another closely related analogue is the dimensionless Love
number in geophysics (e.g. Jeffreys 1970), which measures the ratio of the
direct tidal potential from the Moon to the augmented tidal potential that
includes the gravitational potential arising from the deformation of the Earth
in response to the lunar tide.

The influence of the halo response on tidal fields was first pointed out by
Lynden-Bell (1985), who computed the Love number for a spherical fluid halo
with power-law density distribution, $\rho_0(r)\propto r^{-\alpha}$ (see also
Nelson \& Tremaine 1995); we shall find below, however, that the response of
fluid and stellar systems with the same density distribution can be quite
different. Weinberg (1995, 1997) found that the response of the Galactic halo
strongly enhances the direct tidal field from the Magellanic Clouds and
suggested that the resulting disk distortion could account for the location,
position angle, and sign of the HI warp in the outer Galaxy. In many respects
Weinberg's calculations are much more sophisticated than ours, as we shall
focus on a simplified case that provides insight into the phenomenon rather
than accurate numbers for a realistic system.

In the present paper, we examine the response of a spherical, scale-free
isothermal stellar system to an adiabatically applied external gravitational
perturbation.  This problem only approximates reality in that it neglects the
time-dependence that accompanies most tidal fields (e.g. those due to orbiting
satellites). The reward is an analytically tractable problem that permits us
to understand the physics of the halo response---and the linear response of
stellar systems in general---in a more detailed manner than has hitherto been
possible. Moreover, we are most interested in the response of the galaxy at
small radii, where the dynamical time is short and the approximation of a
static tide is not unreasonable.

Adiabatically applied gravitational perturbations are also relevant to other
astrophysical phenomena, including the slow growth of a central black hole in a
galaxy (e.g. Young 1980, Quinlan et al. 1995) and the slow growth of a galaxy
disk in a halo or spheroid (Binney \& May 1986; Dubinski 1994). Thus we shall
also apply the tools we have developed to these problems. 

The plan of the paper is as follows.  In \S\ref{sec:derivation}, we develop
the solution to the linearized Boltzmann-Poisson equation that describes the
response of a stellar system to an adiabatically applied perturbation.  The
relationship between adiabaticity and reversibility is discussed and a
comparison with analogous fluid systems is also made.  In
\S\ref{sec:iso_sphere}, we derive the adiabatic Green's function for the
response of the isothermal sphere.  This machinery is applied in
\S\ref{sec:apps} and the results are discussed in \S\ref{sec:disc}.

\section{Linearized Adiabatic Response}
\label{sec:derivation}

\noindent
We examine the response of a stationary stellar system with unperturbed
distribution function (hereafter DF) $F(\bfr,\bfv)$ and gravitational
potential $\Phi_0(\bfr)$ to a weak external potential $\Phi^e(\bfr,t)$. The
evolution of the perturbed DF $f(\bfr,\bfv,t)$ is described by the linearized
collisionless Boltzmann equation (e.g. Kalnajs 1971; Weinberg 1989), \be
\label{eq:LBE}
\LBEreal;
\ee
the first equality defines the convective or Lagrangian derivative in phase
space. The perturbing potential
$\Phi_1(\bfr,t)=\Phi^e(\bfr,t)+\Phi^s(\bfr,t),$ where $\Phi^s$ is the
response potential, given by Poisson's equation
\be
\label{eq:poisson}
\nabla^2\Phi^s=4\pi G\int f d\bfv.
\ee

We shall restrict ourselves to the case where the DF of the
stationary stellar system depends only on energy, $F(\bfr,\bfv)=F(E)$,
where $E=\half\bfv^2+\Phi_0(\bfr)$ is the energy per unit mass. Then equation
(\ref{eq:LBE}) becomes 
\be
\label{eq:LLBE1}
\drv{f}{t}=\bfv\cdot\bnabla\Phi_1\drv{F}{E}=F_E\left(
\drv{\Phi_1}{t}-\pdrv{\Phi_1}{t}\right), 
\ee
where $F_E\equiv dF/dE$. 
Note that $-\bfv\cdot\bnabla\Phi_1$ is the power per unit mass delivered by
the perturbing potential. 

In this paper we shall focus on the case where the external perturbation grows
slowly from zero in the distant past. We assume that the potential
$\Phi_0(\bfr)$ admits action-angle
variables $(\bfI,\bfw)$, so that the unperturbed stellar orbits
have the form $\bfI=$constant, $\bfw=\bfomega(\bfI) t+$constant. We then
expand the perturbing potential in a Fourier series of the form
$\Phi_1=\sum_\bfk\Phi_\bfk(\bfI,t)\*\exp(i\bfk\cdot\bfw)$, where $\bfk$ is an
integer triplet. We assume that
$\Phi_\bfk(\bfI,t)=\Phi_\bfk(\bfI)\exp(\epsilon t)$ where $\epsilon$ is small
and positive. Then we can integrate (\ref{eq:LLBE1}) to obtain 
\be
f=F_E\sum_{\bfk}\Phi_\bfk(\bfI,t) \left[\bfk\cdot\bfomega\over
\bfk\cdot\bfomega-i\epsilon\right]\exp(i\bfk\cdot\bfw). 
\label{eq:epseq} 
\ee 
In the limit $\epsilon\to0$ the square bracket approaches unity, unless
$\bfk\cdot\bfomega=0$, in which case it is zero. The condition
$\bfk\cdot\bfomega=0$ can be satisfied in three different ways:

\begin{enumerate}

\item At isolated or local resonances in phase space, where
$\bfk\cdot\bfomega(\bfI)=0$ for a particular set of actions $\bfI$
(accidental degeneracies). We shall ignore such resonances since they
are unlikely to dominate the overall response of the stellar system.

\item When $\bfk={\bf 0}$.

\item When some symmetry of the stellar system dictates that there are
non-zero values of $\bfk$ such that $\bfk\cdot\bfomega(\bfI)=0$ for all values
of the actions $\bfI$ (global resonance). For example, with a suitable choice
of actions, in spherical potentials $\omega_3=0$ (the orbital plane does not
precess) so that $\bfk\cdot\bfomega=0$ whenever $k_1=k_2=0$; and for Kepler
potentials $\omega_1=\omega_2$ so that $\bfk\cdot\bfomega=0$ whenever
$k_1=-k_2$.

\end{enumerate}

For a given stellar system, we shall denote by $N$ the set of integer triples
$\bfk$ such that $\bfk\cdot\bfomega(\bfI)=0$ for all values of the actions
$\bfI$ (i.e. all $\bfk$ satisfying conditions 2 or 3 above).
Then in the adiabatic limit $\epsilon\to0$ equation (\ref{eq:epseq}) becomes
(e.g. Lynden-Bell 1969) \be
\label{eq:adb_case}
f=F_E\sum_{\bfk\not\in N}\Phi_\bfk(\bfI,t)\exp(i\bfk\cdot\bfw)
=F_E\bigl[\Phi_1-\Phiav\bigr], 
\ee 
where $\Phiav=\sum_{\bfk\in N}\Phi_{\bfk}(\bfI,t)\exp(i\bfk\cdot\bfw)$.
$\Phiav$ can be regarded as a time average of the perturbing potential, over
times long compared to the orbital period but short compared to
$\epsilon^{-1}$. This is not the same as an average over orbital phase,
because the time average of $\exp(i\bfk\cdot\bfw)\not=0$ if $\bfk\in N$ but
the phase average of  $\exp(i\bfk\cdot\bfw)\not=0$ only if $\bfk={\bf 0}$. 

Note that the solution (\ref{eq:adb_case}) is only one of many possible static
solutions to the linearized collisionless Boltzmann equation if the set $N$ is
not empty. If we write the perturbed DF in action-angle variables,
$f=\sum_\bfk f_\bfk(\bfI)\exp(i\bfk\cdot\bfw)$, then the linearized
collisionless Boltzmann equation reads
\be
\sum_{\bfk}\exp(i\bfk\cdot\bfw)(\bfk\cdot\bfomega)(f_\bfk-F_E\Phi_\bfk)=0,
\ee
which has the general solution
\be
f=F_E\Phi_1+\sum_{\bfk\in N}p_\bfk(\bfI)\exp(i\bfk\cdot\bfw)
\label{eq:general}
\ee 
where $p_\bfk(\bfI)$ is an arbitrary function; equation
(\ref{eq:adb_case}) corresponds to the particular case $p_\bfk=-F_E\Phi_\bfk$.

	The absence of the term $\bfk={\bf 0}$ in equation
(\ref{eq:adb_case}) has a simple physical interpretation in terms of entropy.
The entropy of the unperturbed stellar system is
\begin{equation}
\label{eq:S_i}
S_i=\int d{\bf r}d{\bf v} F\ln F=(2\pi)^3\int d{\bf I} F\ln F,
\end{equation}
because $F(E)$ is a function of the actions only and $d\bfr d\bfv=d\bfI d\bfw$.
The final entropy is
\begin{equation}
\label{eq:S_f}
S_f=\int d\bfI d\bfw (F+f)\ln(F+f)=S_i+
\int d\bfI d\bfw(1+\ln F)f+\hbox{O}(f^2).
\end{equation}
Using equation (\ref{eq:adb_case}), we find the change in entropy 
\begin{equation}
\label{eq:delta_S_stel}
\Delta S\equiv S_f-S_i=\int d{\bf I}d{\bf w} (1+\ln F)F_E\sum_{\bfk\not\in N}
	\Phi_{\bf k}\exp(i {\bf k}\cdot{\bf w}),
\end{equation}
which vanishes upon integrating over $\bfw$, because ${\bf 0}\in N$. 

We can also compare the solution (\ref{eq:adb_case}) to the response
of a barotropic fluid system to a weak, slowly growing external
potential. In an equilibrium barotropic fluid, the pressure $P$,
density $\rho$ and potential $\Phi$ form a one-parameter family under
an isentropic change; thus
$P=P(\rho)=P(\Phi)$.  The equation of hydrostatic equilibrium states
that $dP=-\rho d\Phi$ so

\begin{equation}
\rho=-{dP\over d\Phi}.
\end{equation}
For a small change in potential $\Phi_1$, we can determine the perturbed
density by a Taylor expansion of this equation. Defining $\rho=\rho_0+\rho_1$,
we find that
\begin{equation}
\rho_1=-\left({d^2P\over d\Phi^2}\right)_0\Phi_1=
\left({d\rho\over d\Phi}\right)_0\Phi_1.
\label{eq:wwwww}
\end{equation}
For comparison to stellar systems, take equation (\ref{eq:general}) with
$p_\bfk=0$, and integrate over velocity: 
\be 
\rho_1=\int f d\bfv=\Phi_1\int
F_E d\bfv.
\ee
Since the unperturbed density $\rho_0=\int F(\half v^2+\Phi_0)d\bfv$, we have
$(d\rho/d\Phi)_0=\int F_E d\bfv$, so that
\be
\rho_1=\left({d\rho\over d\Phi}\right)_0\Phi_1.
\label{eq:xxxfff}
\ee The result (\ref{eq:xxxfff}) is equivalent to (\ref{eq:wwwww}). In other
words the response of a barotropic fluid system to a slowly growing potential
is the same as the static solution 
\be 
f_1=F_E\Phi_1
\label{eq:adf} 
\ee
of the stellar system with the same density distribution, but is {\em not} the
same as the response of this stellar system to the same slowly growing
potential.

In the following sections we shall examine the adiabatic response of both
stellar (eq. \ref{eq:adb_case}) and fluid (eq. \ref{eq:adf}) systems; the
fluid is less realistic but captures much of the relevant physics with less
algebraic complexity. 

\section{Adiabatic response of an isothermal sphere}
\label{sec:iso_sphere}

\noindent
We now investigate the solutions to equations (\ref{eq:poisson}) and 
(\ref{eq:adb_case}) in the case where the stationary stellar system is a
singular isothermal sphere (Chandrasekhar 1939, Binney \& Tremaine 1987), in
which the DF $F(E)$, the density $\rho_0(r)=\int
F d\bfv$, and the potential $\Phi_0(r)$ have the form
\be
F(E)=K\exp(-\beta E),\qquad \rho_0(r)={\rho_ar_a^2\over r^2},\qquad
\Phi_0(r)={2\over\beta}\log(r/r_a),
\label{eq:isodef}
\ee
where $2\pi G\rho_ar_a^2\beta=1$, $K=(2\pi)^{-3/2}\rho_a\beta^{3/2}$.
Equations (\ref{eq:poisson}) and (\ref{eq:adb_case}) can be written as 
\be
\label{eq:iso_Pois}
\nabla^2\Phi^s+{2\over r^2}(\Phi^s+\Phi^e)=-4\pi G\int d\bfv
{F_E}\langle\Phi^s+\Phi^e\rangle_{\bfr,\bfv},
\ee
where $\langle\cdot\rangle_{\bfr,\bfv}$ denotes an
average---in the sense of equation (\ref{eq:adb_case})---over the orbit
passing through $(\bfr,\bfv)$.  This constitutes an integro-differential
equation for the response potential $\Phi^s$. In the case of a fluid the right
side of equation (\ref{eq:iso_Pois}) would be zero.

Write 
\be 
\Phi(\bfr)=\sum_\lm\Phi_\lm(r)Y_\lm(\bfOmega).  
\ee 
Then after multiplying (\ref{eq:iso_Pois}) by $Y_\lm^\ast(\bfOmega)$ and 
integrating over $d\bfOmega$ we get
\begin{eqnarray}
\lefteqn{{1\over r^2}{d\over dr}r^2{d\over dr}\Phi^s_\lm + {1\over
r^2}[2-\ell(\ell+1)]\Phi^s_\lm+{2\over r^2}\Phi^e_\lm}
\qquad\qquad & &\nonumber\\ &=&-4\pi G\int d\bfOmega
Y_\lm^\ast(\bfOmega)\int d\bfv
F_E\left\langle\sum_{\ell'm'}\left(\Phi^s_{\ell'm'}+\Phi^e_{\ell'm'}\right)
Y_{\ell'm'}\right\rangle_{\bfr,\bfv}.
\label{eq:two}
\end{eqnarray}
We now multiply (\ref{eq:two}) by $r^{\alpha+1}$ and integrate over $r$. We
assume that $\Phi^s_\lm\sim r^{-a_-}$ as $r\to 0$ and $\sim r^{-a_+}$ as
$r\to\infty$, where $a_+\le\ell+1$, $a_-\ge-\ell$ since otherwise the density
distribution that gives rise to $\Phi^s_{\ell m}$ is unphysical. 
We restrict $\alpha$ to the strip of the complex plane defined by 
\be 
-\ell\le a_-<\Re(\alpha)<a_+\le\ell+1,
\label{eq:bdef}
\ee
so that boundary terms arising from integration by parts vanish. 

The Mellin transform of a function $y(r)$ is written $\widetilde y(\alpha)$,
where
\be
\widetilde y(\alpha)=\int_0^\infty r^{\alpha-1}y(r)dr,\qquad y(r)={1\over 2\pi
i}\int_{a-i\infty}^{a+i\infty}\widetilde y(\alpha)r^{-\alpha}d\alpha.
\ee
Thus
\begin{eqnarray}
\lefteqn{\alpha(\alpha-1)\tPhi^s_\lm(\alpha)
+[2-\ell(\ell+1)]\tPhi^s_\lm(\alpha)+
2\tPhi^e_\lm(\alpha)}\qquad\qquad\qquad\qquad& &\nonumber \\
&=&-4\pi G \int d\bfr d\bfv
r^{\alpha-1}Y_\lm^\ast(\bfOmega)F_E
\left\langle\sum_{\ell'm'}(\Phi^s_{\ell'm'}+
\Phi^e_{\ell'm'})Y_{\ell'm'}\right\rangle_{\bfr,\bfv}.
\label{eq:three}
\end{eqnarray}

We now convert to action-angle variables $(\bfI,\bfw)$. Our
conventions follow Tremaine \& Weinberg (1984): $I_2$ is the total
angular momentum, $I_3$ is the $z$-component of the angular momentum, and
$I_1$ is the radial action, 
\be
I_1={1\over \pi}\int_{r_p}^{r_a}v_rdr,
\ee
where $v_r=\big[2E-2\Phi_0(r)-I_2^2/r^2\big]^{1/2}$ is the radial velocity and
$r_p$ and $r_a$ are the pericenter and apocenter distances, at which
$v_r=0$. The only angle with a direct geometrical interpretation is $w_3$,
which is the azimuth at which the orbit crosses the equatorial plane upward
(the ascending node).  Motion in the unperturbed Hamiltonian
$H_0(\bfI,\bfw)=\half v^2+\Phi_0(r)$ is given by $\bfI=$constant,
$\bfw=\bfomega t+\bfw_0$ where $\bfomega=\partial H_0/\partial\bfI$. Note that
$\omega_3=0$, since the orbital plane does not precess in a spherical
potential, so that $\bfk\cdot\bfomega=0$ if $k_1=k_2=0$.  Note also that
$d\bfr d\bfv=d\bfI d\bfw$.

Expanding the potentials in action-angle variables, we have 
\begin{eqnarray}
[\Phi^s_{\ell'm'}(r')+\Phi^e_{\ell'm'}(r')]Y_{\ell'm'}(\bfOmega') &=& 
\sum_{l_1'l_2'l_3'}\delta_{l_3'm'}
V_{\ell'l_2'l_3'}(\beta')W^{l_1'}_{\ell'l_2'l_3'}(\bfI')
\exp\left(i\sum_{k=1}^3l_k'w_k'\right),\nonumber\\
r^{\alpha^\ast-1}Y_{\ell m }(\bfOmega ) &=& 
\sum_{l_1 l_2 l_3 }\delta_{l_3m}V_{\ell l_2l_3}(\beta)U^{l_1 }_{\ell l_2}(\bfI)
\exp\left(i\sum_{k=1}^3l_k w_k\right).
\label{eq:actang}
\end{eqnarray}
Here $\cos\beta=I_3/I_2$, $V_{\ell l_2 l_3}(\beta)$ is defined in terms of
rotation matrices by Tremaine \& Weinberg (1984), and
\begin{eqnarray}
W^{l_1}_{\ell l_2 m}(\bfI)&=&{1\over 2\pi}
\int_{-\pi}^\pi dw_1\exp(-il_1w_1)
\big[\Phi^s_{\ell m}(r)+\Phi^e_{\ell m}(r)\big]\exp[-il_2\chi(\bfI,
w_1)],\nonumber\\ 
U^{l_1}_{\ell l_2}(\bfI)&=&{1\over 2\pi}
\int_{-\pi}^\pi dw_1\exp(-il_1w_1)
r^{\alpha^\ast-1}\exp[-il_2\chi(\bfI,w_1)].
\end{eqnarray}
In this equation 
\be
\chi(\bfI,w_1)=\int_{r_p(\bfI)}^{r(\bfI,w_1)}{dr\over v_r}(\omega_2-I_2/r^2).
\ee

We now find the average $\langle\cdot\rangle_{\bfr,\bfv}$ of the first
of equations (\ref{eq:actang}). This is obtained by replacing $\bfI'$
by $\bfI$, $\beta'$ by $\beta$, $w_k'$ by $w_k$ and restricting the
summation to $(l_1',l_2',l_3')\in N$. Now substitute this result into
equation (\ref{eq:three}), replacing $d\bfr d\bfv$ by $d\bfI d\bfw$
and $r^{\alpha-1}Y_{\ell m}(\bfOmega)$ by the complex conjugate of the
second of equations (\ref{eq:actang}). Only terms with $l_1=l_1'$,
$l_2=l_2'$, and $l_3=l_3'=m=m'$ survive the integration over $\bfw$.
Moreover, in the singular isothermal sphere the set of resonant
triplets $N$ is given by $l_1=l_2=0$, since the only global resonance is due
to spherical symmetry. Thus, the right side of (\ref{eq:three}) becomes
\begin{eqnarray}
\lefteqn{-4\pi G \int d\bfr d\bfv
r^{\alpha-1}Y_\lm^\ast(\bfOmega)F_E
\left\langle\sum_{\ell'm'}(\Phi^s_{\ell'm'}+
\Phi^e_{\ell'm'})Y_{\ell'm'}\right\rangle_{\bfr,\bfv}}
\qquad\qquad & & \nonumber\\ 
& = & -2^5\pi^4 G \sum_{\ell'}\int dI_1I_2dI_2d\cos\beta
V^\ast_{\ell 0 m}(\beta)
U^{0\ast}_{\ell 0}(\bfI) F_E V_{\ell' 0 m}(\beta)W^0_{\ell'0m}(\bfI).
\end{eqnarray}

We next use the orthogonality relation (Edmonds 1960)
\be
\int d\cos\beta V^\ast_{\ell l_2 m}(\beta)V_{\ell' l_2 m}(\beta)={2\over
2\ell+1}\left|Y_{\ell l_2}(\half\pi,0)\right|^2\delta_{\ell\ell'}\equiv
C_{\ell l_2}\delta_{\ell\ell'};
\ee
in particular
\begin{eqnarray}
C_{\ell 0} & = & {1\over 2\pi^2} \left[\Gamma(\half\ell+\half)
\over\Gamma(\half\ell+1)\right]^2,\qquad \hbox{$\ell$ even},\nonumber\\
& = & 0,\qquad\qquad\qquad\qquad\quad\, \hbox{$\ell$ odd}.
\end{eqnarray} 
Thus equation (\ref{eq:three}) becomes
\begin{eqnarray}
\lefteqn{
\alpha(\alpha-1)\tPhi^s_\lm(\alpha)
+[2-\ell(\ell+1)]\tPhi^s_\lm(\alpha)+
2\tPhi^e_\lm(\alpha)}& & \nonumber \\
&=&-2^5\pi^4 C_{\ell 0} G \int dI_1I_2dI_2 F_E
U^{0\ast}_{\ell 0}(\bfI) W^0_{\ell 0 m}(\bfI)\nonumber\\
&=&-8\pi^2 C_{\ell 0} G \int dI_1I_2dI_2 F_E \int dw_1r^{\alpha-1}(\bfI,w_1)
\int dw_1'\left(\Phi_\lm^s+\Phi_\lm^e\right)[r(\bfI,w_1')]\nonumber \\
&=&-8\pi^2 C_{\ell 0} G \int {RdRdE I_c^2(E)F_E\over\omega_1(E,R)}
\int dw_1r^{\alpha-1}(\bfI,w_1)
\int dw_1'\left(\Phi_\lm^s+\Phi_\lm^e\right)[r(\bfI,w_1')]; \nonumber \\
&& 
\label{eq:five}
\end{eqnarray}
in the last line we have changed the integration variables from $I_1,I_2$ to
the energy $E$ and the dimensionless angular momentum $R\equiv I_2/I_c(E)$,
where $I_c(E)$ is the angular momentum of a circular orbit of energy $E$.

Since the potential is scale-free the radius can be written in the
form $r(\bfI,w_1)=r_c(E)\*x(R,w_1)$, where $x$ is a dimensionless function that
can be determined by numerical integration of the orbits. Changing the
integration variable from $E$ to $r'=r_c(E)x(w_1',R)$, the 
right side of (\ref{eq:five}) becomes
\begin{eqnarray}
\lefteqn{-8\pi^2 C_{\ell 0} G \int RdR\int dw_1x^{\alpha-1}(R,w_1)}
\qquad\qquad & & \nonumber\\
& \times & \int dw_1'
x^{-\alpha}(R,w_1')\int dr'{r'}^{\alpha-1}\left[\Phi_\lm^s(r')+\Phi_\lm^e(r')
\right]{dE\over dr_c}{I_c^2(E)F_E\over \omega_1(E,R)},
\end{eqnarray}
For the singular isothermal sphere
\be
{dE\over dr_c}{I_c^2(E)F_E\over \omega_1(E,R)}=-{1\over
2^{1/2}\pi^{5/2}eGg(R)},\qquad\hbox{where }\qquad
g(R)\equiv \omega_1(E,R)r_c(E)\beta^{1/2}
\ee
is a function only of $R$. Thus
\begin{eqnarray}
\lefteqn{
\alpha(\alpha-1)\tPhi^s_\lm(\alpha)
+[2-\ell(\ell+1)]\tPhi^s_\lm(\alpha)+
2\tPhi^e_\lm(\alpha)}\qquad\qquad & & \nonumber  \\
&=&{2^{5/2} C_{\ell 0}\over \pi^{1/2}e} \int {RdR\over g(R)}
\int dw_1x^{\alpha-1}(R,w_1)\int dw_1'x^{-\alpha}(R,w_1')
\left[\tPhi_\lm^s(\alpha)+\tPhi_\lm^e(\alpha)
\right]\nonumber \\
& \equiv & C_{\ell 0}H(\alpha)
\left[\tPhi_\lm^s(\alpha)+\tPhi_\lm^e(\alpha)\right],
\label{eq:master}
\end{eqnarray}
where
\be
H(\alpha)={2^{5/2}\over \pi^{1/2}e}\int {RdR\over g(R)}
\int dw_1x^{\alpha-1}(R,w_1)\int dw_1'
x^{-\alpha}(R,w_1').
\label{eq:hdef}
\ee
Note that
\be
H^\ast(\alpha)=H(\alpha^\ast),\qquad H(1-\alpha)=H(\alpha).
\label{eq:hsymm}
\ee
It can also be shown after some algebra that
\be
H(0)=H(1)=4\pi; \qquad H'(1)=-H'(0)=\pi.
\label{eq:htriv}
\ee

Equation (\ref{eq:master}) can be rewritten as
\be
\tPhi^s_\lm(\alpha)={C_{\ell 0}H(\alpha)-2\over
D_\ell(\alpha)}\,\tPhi^e_\lm(\alpha),
\label{eq:sssss}
\ee
where
\begin{equation}
\label{eq:denom}
D_\ell(\alpha)\equiv\alpha(\alpha-1)+[2-\llp]-C_{\ell 0}H(\alpha).
\end{equation}
Both $H(\alpha)$ and $D_\ell(\alpha)$ are even functions of $\alpha-\half$.

Thus the linear response of the singular isothermal sphere to adiabatic
perturbations for odd $\ell$ is analytic and for even $\ell$ requires only the
numerical evaluation of a single function $H(\alpha)$.

\subsection{Zero-frequency normal modes}

\noindent
Roots of the dispersion relation $D_\ell(\alpha)=0$ may be regarded as
zero-frequency normal modes of the singular isothermal sphere.  The
contribution to $\Phi^s_\lm(r)$ from each root of $D_\ell(\alpha)$ is
proportional to $r^{-\alpha}$, which follows from the inverse Mellin
transform. \omit{(Of course, there are other zero-frequency modes that do not
satisfy this dispersion relation [cf. eq. \ref{eq:general}], but these are not
excited by an adiabatically growing external potential.)} In general these
solutions should not be interpreted as physical normal modes, since a
power-law potential perturbation is always nonlinear at large or small radii,
except in the special case $\Re(\alpha)=0$. In the fluid case, this limitation
is highlighted by Lichtenstein's theorem, which states that static, isolated,
equilibrium fluid systems {\it must} be spherical, i.e. there are no finite
normal modes with zero frequency (e.g. Lindblom 1992).

Two trivial normal modes arise from gauge transformations: the mode
$\alpha=0$, $\ell=0$ corresponds to a shift in the zero-point of the
potential, and $\alpha=1$, $\ell=1$ corresponds to a uniform translation of
the unperturbed stellar system.

The symmetries (\ref{eq:hsymm}) imply that if $\alpha$ is a root of the
dispersion relation, then so is $\alpha^\ast$ and $1-\alpha$. Thus if $x\equiv
\alpha-\half$ the roots come in pairs $\half+x$, $\half-x$ when $x$ is real or
imaginary, and in quartets $\half+x$, $\half-x$, $\half+x^\ast$,
$\half-x^\ast$ if $x$ is complex.

Finding the roots of $D_\ell(\alpha)$ is easy when $\ell$ is odd: then
$C_{\ell 0}=0$ so that the roots are
\be
\alpha=\half\pm\left[\ell(\ell+1)-
\textstyle{7\over4}\right]^{1/2},\qquad\ell\hbox{ odd}.
\label{eq:gasdisp}
\ee
To find the roots when $\ell$ is even, it is convenient to rewrite the
function $H(\alpha)$ explicitly as a complex function of a complex variable.
Defining $\alpha=a+bi$ where $a$ and $b$ are real, we find the real and
imaginary parts of $H(\alpha)$ to be
\begin{eqnarray}
\label{eq:H_a_ri}
\Re[H(\alpha)]& = &{2^{5/2}\over \pi^{1/2}e}
\int{RdR\over g(R)}\int dw_1\int dw_1'
\cos [b\ln (x/x')]x^{a-1}x'^{-a},\nonumber\\
\Im[H(\alpha)]& = &{2^{5/2}\over \pi^{1/2}e}\int{RdR\over g(R)}
\int dw_1\int dw_1'
\sin [b\ln (x/x')]x^{a-1}x'^{-a}.
\end{eqnarray}

The roots of $D_\ell(\alpha)$ occur at the simultaneous zeros of its
real and imaginary parts.  Figure 1 shows the zero curves in the
complex plane of the real and imaginary parts of $D_{\ell}(\alpha)$,
for several values of $\ell$.  The roots with the smallest values of
$|a|=|\Re(\alpha)|$, which we denote $\alpha=a_0^+\pm ib_0$ and
$a_0^-\pm ib_0$, have the greatest physical significance, since these
generally determine the asymptotic behaviour of the response of the
system at small and large radii. These roots are given in Table
\ref{tab:asymp_roots}; note that they satisfy the constraint $-\ell\le
a_-,a_+\le \ell+1$ given by equation (\ref{eq:bdef}).  Also note that
$a_0^-+a_0^+=1$.

The normal mode corresponding to the root $a_0^+=1$ for $\ell=1$ is simply a
uniform displacement of the unperturbed system.

\begin{figure}
\label{fig:roots}
\plotone{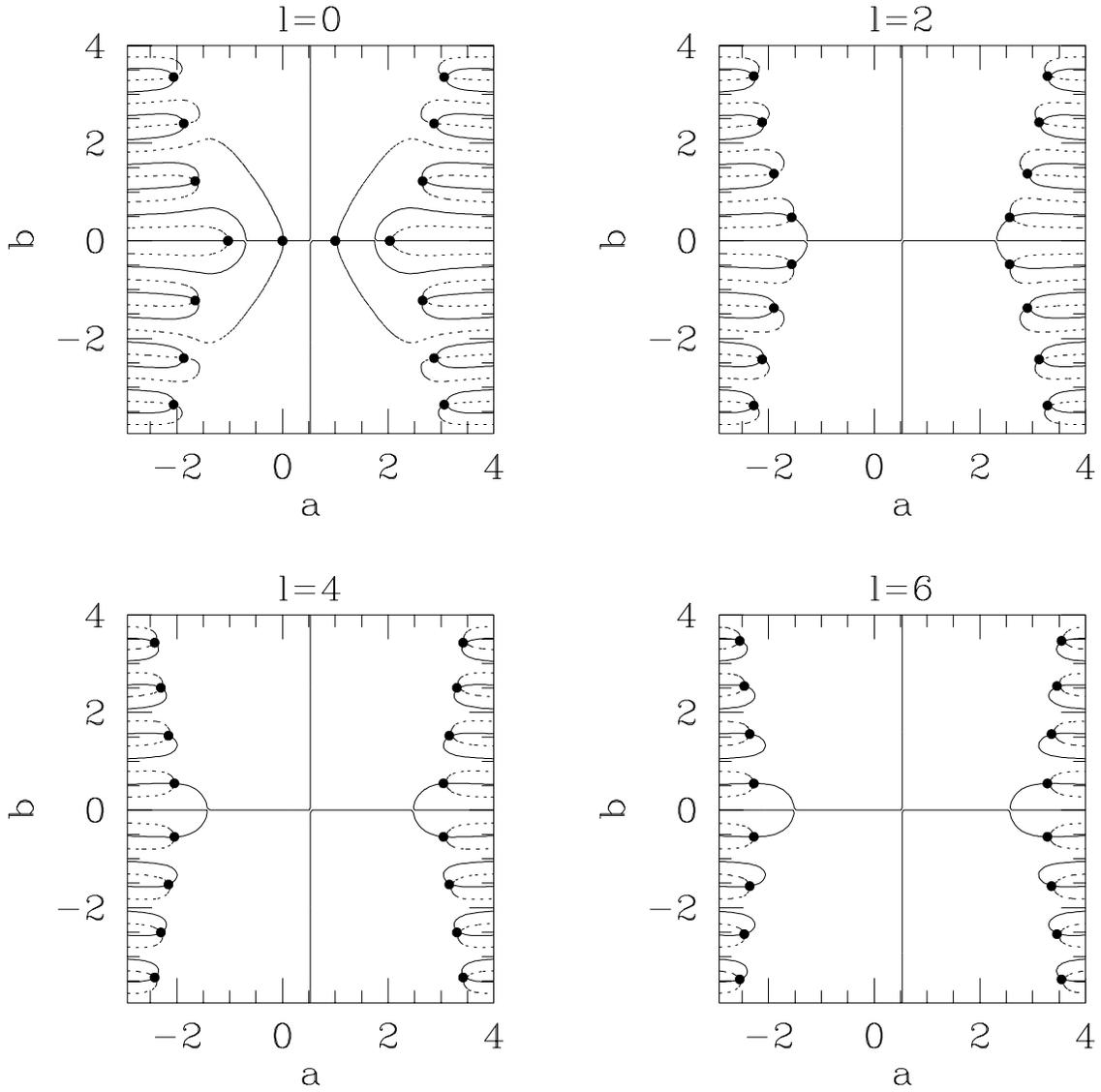}
\caption{The zero curves of the real (dotted) and imaginary (solid)
        parts of $D_\ell(\alpha)$ (eq. \ref{eq:denom}) for indicated
        values of $\ell$.  The roots (solid points) lie at the
        intersections of the two curves.  Those with largest real
        parts in the left half-plane will dominate the response at
        small radii.}
\end{figure}

\begin{table*}
\caption{Asymptotically dominant roots}
\label{tab:asymp_roots}
\begin{tabular}{lrrr}
\\
$\ell$&$a_0^-$&$a_0^+$&$b_0$\\
$0$&$0.0000$&$1.0000$&$0.0000$\\
$1$&$0.0000$&$1.0000$&$0.0000$\\
$2$&$-1.5623$&$2.5623$&$0.4804$\\
$3$&$-2.7016$&$3.7016$&$0.0000$\\
$4$&$-2.0460$&$3.0460$&$0.5480$\\
$5$&$-4.8151$&$5.8151$&$0.0000$\\
$6$&$-2.2771$&$3.2771$&$0.5457$\\
$7$&$-6.8655$&$7.8655$&$0.0000$\\
\end{tabular}
\end{table*}

For comparison, the dispersion relation in the fluid system is obtained by
setting $H(\alpha)=0$ in equation (\ref{eq:denom}). The corresponding roots
are given by (\ref{eq:gasdisp}) for all $\ell$, both even and odd. Thus for
$\ell=0$ the fluid system has $\alpha=\half(1\pm i\sqrt{7})$; this is a
familiar result since the difference in potential between the singular and
non-singular isothermal spheres oscillates with this spatial frequency at
large radii (Chandrasekhar 1939). For $\ell=2$, equation (\ref{eq:gasdisp})
yields $\alpha=\half(1\pm\sqrt{17})$.

\subsection{Green's function}
\label{sec:iso_pm}

\noindent
We now discuss the adiabatic response of the singular isothermal sphere to a
point mass $m_p$ at position $\bfr_p$, which corresponds to the external
potential
\be
\label{eq:pm_exp}
\Phi^e(\bfr)=\sum_\lm\Phi^e_\lm(r)Y_\lm(\bfOmega),\qquad \hbox{where} \qquad
\Phi^e_\lm(r)=-{4\pi Gm_p\over 2\ell +1}
	Y_{\ell m}^*(\bfOmega_p){r_<^{\ell}\over
	r_>^{\ell+1}}. 
\ee
Here $\bfOmega_p$ denotes the angular coordinates of $\bfr_p$,
$r_<=\min(|\bfr|,|\bfr_p|)$ and $r_>=\max(|\bfr|,|\bfr_p|)$.  The Mellin
transform of the point-mass potential is therefore given by 
\be
\label{eq:pm_exp_trans}
\tPhi^e_\lm(\alpha)=-{4\pi Gm_p \over 2\ell +1}
	Y_{\ell m}^*(\bfOmega_p) r_p^{\alpha-1}
	\left({1\over \alpha+\ell}-{1\over \alpha-\ell-1}\right),\qquad -\ell
< \Re(\alpha) < \ell+1.
\ee

We now employ equation (\ref{eq:sssss}) to find the response
\be
\label{eq:iso_resp_trans}
\tPhi^s_\lm(\alpha)=-{4\pi Gm_p\over 2\ell +1}Y_{\ell
	m}^*(\bfOmega_p)r_p^{\alpha-1}
	\left({1\over \alpha+\ell}-{1\over \alpha-\ell-1}\right)
	{C_{\ell 0}H(\alpha)-2\over D_\ell(\alpha)}.
\ee
The inverse Mellin transform is 
\begin{eqnarray}
\Phi^s_\lm(r)& = & {1\over2\pi i}\int_{a-i\infty}^{a+i\infty}
\tPhi^s_\lm(\alpha)r^{-\alpha}d\alpha \nonumber \\
& = & {Gm_p\over r_p}{2i\over 2\ell +1}Y_\lm^*(\bfOmega_p)
\int_{a-i\infty}^{a+i\infty}d\alpha \left(r_p\over r\right)^\alpha
\left({1\over \alpha+\ell}-{1\over \alpha-\ell-1}\right)
	{C_{\ell 0}H(\alpha)-2\over D_\ell(\alpha)}. \nonumber \\
&&
\label{eq:qqqq}
\end{eqnarray}
The constant $a$ must be chosen so that $-\ell<a<\ell+1$ and $a_-<a<a_+$
(eq. \ref{eq:bdef}). If we choose $a=\half$ and use the fact that $H(\alpha)$
and $D_\ell(\alpha)$ are even functions of $\alpha-\half$, then it is easy to
show that
\be
\Phi^s_\lm(r)={1\over(rr_p)^{1/2}}\Psi_\lm\left(r\over r_p\right),
\qquad\hbox{where}\qquad \Psi_\lm(u)=\Psi_\lm(1/u);
\label{eq:symmxx}
\ee
in other words the response for $r<r_p$ is determined by the response for
$r>r_p$.

To evaluate the integral (\ref{eq:qqqq}) in closed form, we analytically
continue the integrand over the whole complex plane and then close the
integration contour, in the left half-plane for $r\leq r_p$ and in the right
half-plane for $r>r_p$.  The contributions to the integral come from the poles
at $\alpha=-\ell$, $\alpha=\ell+1$, and the roots of $D_\ell(\alpha)$, which
we denote by $\alpha=\half\pm x_k$, where $\Re(x_k)>0$. The asymptotic
behavior at small radii, $\Phi_\lm^s\sim r^{-a_-}$, is determined by the pole
in the negative half-plane with the smallest $|\Re(\alpha)|$ (either $a_0^-$
from Table \ref{tab:asymp_roots} or $-\ell$), while the asymptotic behavior at
large radii, $\Phi_\lm^s\sim r^{-a_+}$, is determined by the pole in the
positive half-plane with the smallest $|\Re(\alpha)|$ (either $a_0^+$ from
Table \ref{tab:asymp_roots} or $\ell+1$). Thus
\begin{eqnarray}
\label{eq:iso_phi_s}
\Phi^s_{\ell m}(r)& = &{4\pi Gm_p\over (2\ell+1)r_p}Y_\lm^\ast(\bfOmega_p)
	\left[\left(r\over r_p\right)^{\ell\phantom{+1}}+\sum_{k}
	{2\ell+1\over D_\ell'(\half-x_k)}\left(r\over r_p\right)^{x_k-\half}
	\right],\quad          r\leq r_p, \nonumber \\
        & = &          {4\pi Gm_p\over (2\ell+1)r_p}Y_\lm^\ast(\bfOmega_p)
	\left[\left(r_p\over r\right)^{\ell+1}-\sum_{k}
	{2\ell+1\over D_\ell'(\half+x_k)}\left(r_p\over r\right)^{x_k+\half}
	\right],\quad          r > r_p;
\end{eqnarray}
these two expressions satisfy the symmetry relation (\ref{eq:symmxx}). The sum
is over all roots with $\Re(\alpha-\half)>0$; recall that if
$\Im(\alpha)\not=0$ then $\alpha^\ast$ is also a root.

The first term on the right side of each equality defines a response
that precisely cancels the external perturbing potential $\Phi_{\ell m}^e$ (eq.
\ref{eq:pm_exp}). The total potential perturbation $\Phi_{\ell
m}^t=\Phi_{\ell m}^e+\Phi_{\ell m}^s$ therefore is entirely determined
by the spectrum of normal modes. When $\ell$ is even the roots $x_k$
are generally complex (Table \ref{tab:asymp_roots}) so $\Phi_{\ell
m}^t(r)$ varies sinusoidally in $\log r$; for odd $\ell$ the total
potential has no phase variation because the roots are real.

Since $D_\ell'(\alpha)$ is an odd function of $\alpha-\half$, the response
potential $\Phi^s_{\ell m}(r)$ is continuous through the shell $r=r_p$. 
The total response $\Phi_s(\bfr)=\sum_{\ell,m}\Phi^s_{\ell m}(r)Y_{\ell
m}(\bfOmega)$ is also continuous.  In addition, $d\Phi^s_{\ell m}/dr$ is 
continuous through the
shell $r=r_p$; this can be shown using the identity 
\be
\label{eq:sum1}
\sum_{k} {x_k\over D_\ell'(\half+x_k)}=\half, 
\ee 
which in turn can be derived by considering the integral $\int_Cd\alpha
(\alpha-\half)/D_\ell(\alpha)$ where $C$ is the circle $|\alpha|\to\infty$.

In the case of a fluid system, the analogous expressions are
\begin{eqnarray}
\label{eq:iso_phig_s}
\Phi^s_{\ell m}(r)& = &{4\pi Gm_p\over (2\ell+1)r_p}Y_\lm^\ast(\bfOmega_p)
	\left[\left(r\over r_p\right)^{\ell\phantom{+1}}-
	{2\ell+1\over [4\ell(\ell+1)-7]^{1/2}}\left(r\over
r_p\right)^{[\ell(\ell+1)-7/4]^{1/2}-1/2}
	\right],\quad          r\leq r_p \nonumber \\
        & = &          {4\pi Gm_p\over (2\ell+1)r_p}Y_\lm^\ast(\bfOmega_p)
	\left[\left(r_p\over r\right)^{\ell+1}-
	{2\ell+1\over [4\ell(\ell+1)-7]^{1/2}}\left(r_p\over 
r\right)^{[\ell(\ell+1)-7/4]^{1/2}+1/2}\right],\quad r > r_p. \nonumber\\
&&
\end{eqnarray}
A similar result was derived already by Lynden-Bell (1985), with minor
differences---in particular he assumed that the fluid halo stopped at $r_p$,
which changes $\Phi^t_{\ell m}$ by a multiplicative factor.

Some care is required to interpret these expressions when $\ell=0$ or
$\ell=1$. For $\ell=0$, the response potential interior to the point mass
contains constant terms ($\propto r^\ell$ and $\propto r^{x_k-1/2}$ for
$x_k=\half$); these terms exert no force and can be eliminated by re-defining
the zero-point of the potential. When $\ell=1$, the concept of the linear
adiabatic response to a small perturbing potential is ill-defined, since 
a small but steady perturbing potential can lead to a large shift in the
center of mass of the stellar system. This indeterminacy can be reflected in
equations (\ref{eq:iso_phi_s}) and (\ref{eq:iso_phig_s}) by adding an
arbitrary amount of the normal mode $\Phi^s_{1m}\propto r^{-1}$ that
corresponds to a uniform translation of the unperturbed stellar system. 

In practice, it is easiest to determine the response potential through direct
numerical calculation of the inverse Mellin transform (\ref{eq:qqqq}), rather
than by summing the residues at all the poles.  The simplest choice of contour
is along the symmetry axis, $a=\half$, for which the imaginary part of
$H(\alpha)$ vanishes.  The real part of equation (\ref{eq:H_a_ri}) then
becomes 

\begin{equation}
H(\half+ib)={2^{5/2}\over \pi^{1/2}e}\int{RdR\over g(R)}
\left\{\left[\int dw_1\cos (b\ln x)x^{-1/2}\right]^2+
		\left[\int dw_1\sin(b\ln x)x^{-1/2}\right]^2\right\},
\end{equation}
which has a narrow peak about $b=0$ and decays approximately
as $b^{-1}$ for $b\gta 50$.

It is also helpful to rewrite equation (\ref{eq:qqqq}) as a Fourier transform
which can be evaluated using FFTs (e.g. Acton 1990):
\be
\Phi^s_\lm(r)=-{4Gm_p\over (r_pr)^{1/2}}Y_\lm^*(\bfOmega_p)
\int_0^{\infty}db\cos\left[b\ln (r/r_p)\right]
{C_{\ell 0}H(\half+ib)-2\over [(\half+\ell)^2+b^2]D_\ell(\half+ib)},
\label{eq:fint}
\ee
The integration parameter $b$ represents the logarithmic 
wavenumber of the response. In the fluid case,
\be
\Phi^s_\lm(r)=-{4Gm_p\over (r_pr)^{1/2}}Y_\lm^*(\bfOmega_p)
\int_0^{\infty}db\cos\left[b\ln (r/r_p)\right]
{2\over [(\half+\ell)^2+b^2][(\half+\ell)^2+b^2-2]}.
\label{eq:fintf}
\ee

\subsection{Density response}
\label{sec:density}

\noindent
The density response is determined directly from the Green's
function (\ref{eq:iso_phi_s}) through Poisson's equation.  
Write the density as a multipole expansion:
\begin{equation}
\rho^s(\bfr)=\sum_{\ell m}\rho^s_{\ell m}(r)Y_{\ell m}(\bf\Omega);
\end{equation}
then substitute into Poisson's equation so that
\begin{equation}
4\pi G\rho^s_{\ell m}(r) = {1\over r^2}\drv{}{r}r^2\drv{\Phi^s_{\ell m}}{r}
				-{\llp\over r^2}\Phi^s_{\ell m}.
\end{equation}
There is no surface-density layer at $r=r_p$, because the response potential
and its gradient are continuous at $r=r_p$ (see the discussion following
equation \ref{eq:iso_phi_s}). 

Substituting equation (\ref{eq:iso_phi_s}) for $\Phi^s_{\ell m}$ gives
\begin{eqnarray}
\label{eq:rho_lm}
\rho^s_{\ell m}(r) & = & {m_p\over r_p^3}Y^*_{\ell m}(\bfOmega_p)
        \sum_{k}\left({r\over r_p}\right)^{x_k-5/2}
        {[x_k^2-(\ell+\half)^2] \over D_\ell'(\half-x_k)},\qquad 
         r\leq r_p \nonumber \\
                &=& {m_p\over r_p^3}Y^*_{\ell m}(\bfOmega_p)\sum_{k}
        \left({r_p\over r}\right)^{x_k+5/2}{[x_k^2-(\ell+\half)^2] 
        \over D_\ell'(\half-x_k)},\qquad r > r_p. \nonumber\\
&&
\end{eqnarray}
At small radii, the fractional density
perturbation is $\rho^s_{\ell m}/\rho_0\sim r^{-a_0^-}$, while at
large radii the fractional perturbation varies as $r^{-a_0^+}$
(cf. Table \ref{tab:asymp_roots}).

For a fluid system, 
\begin{eqnarray}
\label{eq:rho_lmf}
\rho^s_{\ell m}(r) & = & {m_p\over r_p^3}{2Y^*_{\ell m}(\bfOmega_p)
\over [4\ell(\ell+1)-7]^{1/2}}
\left({r\over r_p}\right)^{[\ell(\ell+1)-7/4]^{1/2}-5/2},\quad 
         r\leq r_p \nonumber \\
                &=& {m_p\over r_p^3}{2Y^*_{\ell m}(\bfOmega_p)
        \over[4\ell(\ell+1)-7]^{1/2}}
        \left({r_p\over r}\right)^{[\ell(\ell+1)-7/4]^{1/2}+5/2},
\quad r > r_p.\nonumber\\
\end{eqnarray}

\subsection{Amplification factor}
\label{sec:amp}

\noindent	
It is useful to express these results in terms of the overall
amplification of the point-mass perturbation by the response.
Defining the amplification $\chi_{\ell}(r)=(\Phi^s_{\ell
m}+\Phi^e_{\ell m})/\Phi^e_{\ell m}$, we find that
\begin{eqnarray}
\label{eq:chi_s}
\chi_{\ell}(r)& = &-\sum_{k} {2\ell+1\over D_\ell'(\half-x_k)}
\left(r\over r_p\right)^{x_k-\ell-\half},\qquad r\leq r_p \nonumber \\
        & = &\phantom{-}\sum_{k}
{2\ell+1\over D_\ell'(\half+x_k)}\left(r_p\over r\right)^{x_k-\ell-\half},
\qquad r > r_p.
\end{eqnarray}
The amplification also obeys a symmetry relation similar to 
(\ref{eq:symmxx}), $\chi_{\ell}(r)=\psi_{\ell}(r/r_p)$ where
$\psi_{\ell}(u)=\psi_{\ell}(1/u)$. 

As we discussed after equation (\ref{eq:iso_phig_s}) these formulae are not
meaningful for $\ell=0$ when $r<r_p$, or for $\ell=1$. 

Asymptotically, $\chi_{\ell}(r)\sim
r^{x_0-\ell-1/2}=r^{-a_0^--\ell}$ as $r\to 0$, and
$\chi_{\ell}(r)\sim r^{-x_0+\ell+1/2}=r^{-a_0^++\ell+1}$ as $r\to
\infty$. Values of $a_0$ are given in Table \ref{tab:asymp_roots} and
we see that for all $\ell>0$ the amplification diverges as $r\to 0$ or
$r\to\infty$.  However, for even $\ell$ this asymptotic behavior does
not necessarily appear until very large values of $|\log r|$ are
reached.  For odd $\ell$, there is no distinction between the exact
and asymptotic behavior since the response is determined by a single
root.

\section{Applications}
\label{sec:apps}

\subsection{Tidal amplification}
\label{sec:tide}

\noindent
Consider the response of the halo to a point-mass satellite on a circular
orbit of radius $r_p$. Our assumption that the satellite perturbation is
adiabatic is not accurate at radii $r\sim r_p$, where the characteristic
orbital period in the halo is comparable to the satellite's orbital
period. Nevertheless, the adiabatic approximation is plausible at radii $r\ll
r_p$, and should approximately describe how the tidal field of the satellite
is propagated to small radii. The monopole ($\ell=0$) response of the halo is
not so interesting, since it is difficult to distinguish observationally from
the potential of the unperturbed stellar system. As we have discussed, our
formalism is not powerful enough to determine the dipole ($\ell=1$)
response. Thus we shall focus on the quadrupole ($\ell=2$) response.

Figure 2 shows the amplification $\chi_2(r)$ of the quadrupole
tidal field from a point mass at radius $r_p$.  The amplification has a weak
maximum of about 2 at $\log (r/r_p)\approx -1$ and becomes negative for $\log
r/r_p\simless -2$. This oscillation arises because the asymptotically dominant
root in Table \ref{tab:asymp_roots} is complex.  The amplification remains
negative down to $\log r/r_p\approx 5$---far too small to be of
interest---where it begins a positive rise.  The dashed line shows the much
stronger quadrupole amplification for the analogous fluid system.
	
\begin{figure}
\label{fig:amp}
\plotone{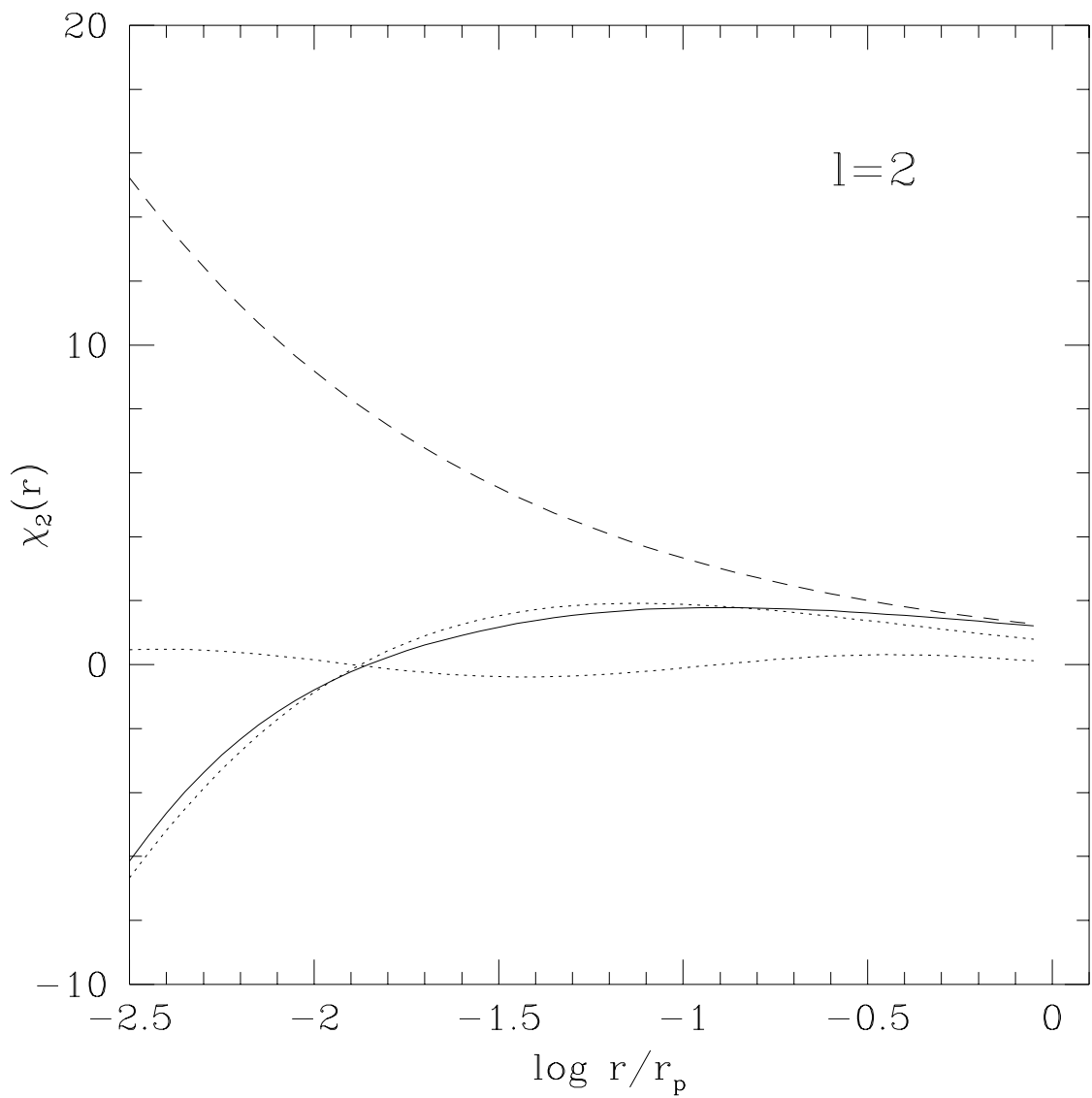}
\caption{Amplification $\chi_2(r)$ of the quadrupole tidal field from
a point mass at $r_p$.  The solid line shows the exact result from
numerical evaluation of equation (\ref{eq:chi_s}).  The dotted line
that follows the solid line shows the amplification due to the
asymptotically dominant root.  The dotted line that oscillates weakly
about zero shows the contribution from the next most important root.
The dashed line shows the much stronger amplification for the fluid
case.}
\end{figure}

These results indicate that the response of an isothermal halo does
not greatly enhance the quadrupole tidal field from an external
point mass: the response is much smaller than for an isothermal fluid
and over a large range in $\log (r/r_p)$ the halo response actually
screens or suppresses the external tidal field.

\subsection{Response to disk growth}
\label{sec:disk}

\noindent
We can also calculate the response of the halo to the adiabatic growth
of an embedded disk (Binney \& May 1986; Dubinski 1994). We compute
the response using scale-free, axisymmetric, razor-thin disk models,
which have density distributions
\begin{equation}
\rho(\bfr)=\Sigma(r){\delta(\cos\theta)\over r}.
\end{equation}
We shall use the relation 
\be 
\delta(\cos\theta)=2\pi\sum_\ell 
Y^*_{2\ell,0}(\half\pi,0) Y_{2\ell,0}(\theta,0);
\ee 
note that only even harmonics contribute to the sum.

Using equation (\ref{eq:iso_phi_s}), we calculate the total potential
arising from the embedded disk:
\begin{eqnarray}
\Phi^t_{2\ell,0}(r) &=& 8\pi^2 G Y^*_{2\ell, 0}\left(\half\pi,0\right)
	\sum_{k}\biggl\{{1\over D'_{2\ell}(\half-x_k)}\int_r^{\infty}dr'
	\left({r\over r'}\right)^{x_k-\half} \Sigma(r') \nonumber \\ 
&&\qquad\qquad\qquad\qquad\qquad-{1\over D'_{2\ell}(\half+x_k)}\int_0^{r}dr'
	\left({r'\over r}\right)^{x_k+\half} \Sigma(r')\biggr\}.
\end{eqnarray}

For example, in the case of a Mestel disk, $\Sigma(r)=\Sigma_ar_a/r$, the
monopole potential is
\begin{equation}
\Phi^t_{00}(r)={4\pi^{3/2}G\Sigma_a r_a\over D'_0(1)} \ln r;
\end{equation}
here we have neglected constant contributions to the potential
(including a divergent one) which change the zero-point but have no
physical effects. For $\ell>0$, we have
\begin{equation}
\Phi^t_{2\ell,0}(r)=16\pi^2G\Sigma_a r_a 
	Y^*_{2\ell,0}\left(\half\pi,0\right)
	\sum_{k} {x_k\over D'_{2\ell}(\half+x_k)({1\over 4}-x_k^2)}.
\label{eq:phidisk}
\end{equation}
These terms give rise to tangential forces but no radial forces.

We can evaluate sums over roots by analogy with equation (\ref{eq:sum1}), using
the integral $\int_C d\alpha
(\alpha-\half)/D_{\ell}(\alpha)(\alpha+1-\gamma)(\alpha-2+\gamma)$.  This
leads to the identity
\begin{equation}
\sum_{k} {x_k\over D'_{\ell}(\half-x_k)[({3\over 2}-\gamma)^2-x_k^2]}=
	{1\over 2D_{\ell}(2-\gamma)}.
\end{equation}
Using this result for $\gamma=1$ and the identity (\ref{eq:htriv}), equation
(\ref{eq:phidisk}) simplifies to
\begin{equation}
\Phi^t_{2\ell,0}(r)={8\pi^2G\Sigma_a r_a 
	Y^*_{2\ell,0}\left(\pi/2,0\right)
	\over 2-2\ell(2\ell+1)-4\pi C_{2\ell,0}},
\label{eq:phidiska}
\end{equation}
which, remarkably, is completely analytic. 

We can easily extend these results to disks with surface density
$\Sigma(r)=\Sigma_a(r_a/r)^\gamma$, $1<\gamma<2$. The total potential is
\begin{equation}
\Phi^t_{2\ell,0}(r)={8\pi^2G\Sigma_ar_a\over D_{2\ell}(2-\gamma)}
         \left(r_a\over r\right)^{\gamma-1} 
	Y^*_{2\ell,0}\left(\half\pi,0\right).
\label{eq:phidiskb}
\end{equation}
We can also express this result in terms of the amplification
$\chi_{2\ell}=\Phi^t_{2\ell,0}/\Phi^d_{2\ell,0}$, where $\Phi^d$ is the
direct potential from the disk. We have 
\be 
\chi_{2\ell}={(1-\gamma)(2-\gamma)-2\ell(2\ell+1)\over
(1-\gamma)(2-\gamma)-2\ell(2\ell+1)+2-C_{2\ell}H(2-\gamma)}.  
\ee 
This expression is valid for all values of $2\ell$ and $\gamma\in[1,2]$ except
for $2\ell=0,\gamma=1$. Note that for large $\ell$, $\chi_{2\ell}\rightarrow
1$; the halo does not amplify the response from the disk on small scales.
Table \ref{tab:disk_chi} shows values of $\chi_{2\ell,0}$ for a range of
$2\ell$ and $\gamma$.

\begin{table*}[ht]
\caption{Disk amplification}
\label{tab:disk_chi}
\begin{tabular}{lrrrr}
\\
$2\ell$/$\gamma$&$1$&${5\over 4}$&${3\over 2}$&${7\over 4}$\\
$0$&2.00&1.93&1.90&1.93\\
$2$&1.39&1.33&1.33&1.33\\
$4$&1.10&1.10&1.10&1.10\\
$6$&1.05&1.05&1.05&1.05\\
$8$&1.03&1.03&1.03&1.03\\
\end{tabular}
\end{table*}

Using Poisson's equation, we derive the response density of
the halo in the presence of the disk.  The response consists
of a density
\be
\rho^s={2\pi\Sigma_ar_a^\gamma\over r^{\gamma+1}}
	\sum_{\ell}(\chi_{2\ell}-1)Y^*_{2\ell, 0}(\half\pi,0)
	Y_{2\ell, 0}(\theta,0);
\ee
This can be expressed as an induced enhancement $\delta$ in
the local halo density
\be
\delta\equiv{\rho^s\over \rho_0}={2\pi\Sigma_ar_a^{\gamma-2}\over
\rho_ar^{\gamma-1}}\sum_{\ell}
	(\chi_{2\ell}-1)Y^*_{2\ell, 0}(\half\pi,0)Y_{2\ell, 0}(\theta,0),
\label{eq:deldef}
\ee
where $\rho_a$ is defined in equation (\ref{eq:isodef}). Figure 3 shows
contours of the fractional density enhancement $\delta$ for various values of
$\gamma$. 

\begin{figure}
\label{fig:rho_gamma}
\plotone{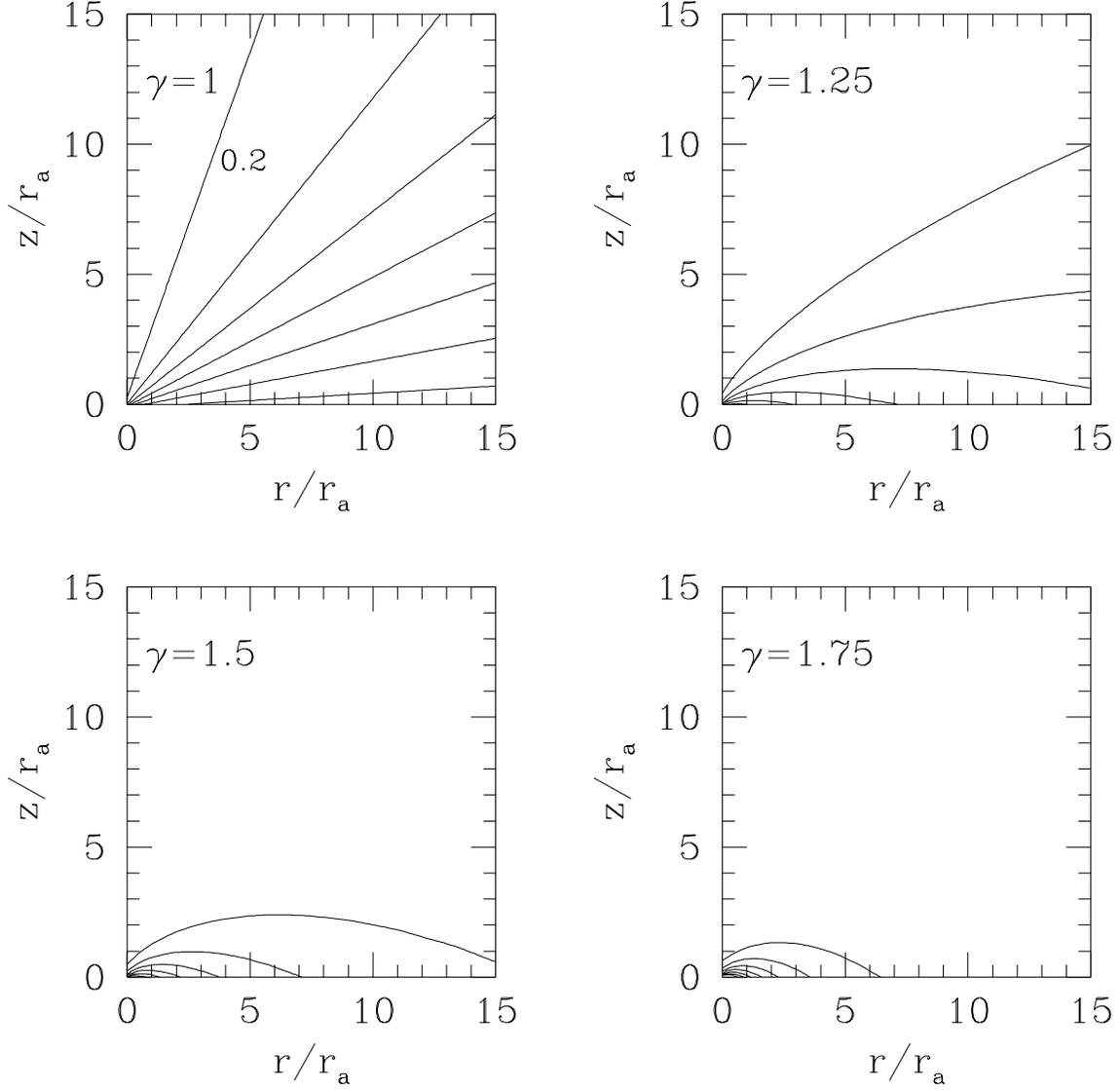}
\caption{Contours of the fractional halo response density $\delta$
(eq. \ref{eq:deldef}) induced by razor-thin disks with surface density
$\Sigma_a(r_a/r)^\gamma$.  Contours show levels $\delta=0.2,0.3,\ldots,0.8$ in
units of $\Sigma_a/r_a\rho_a$, starting from the topmost contour in each panel
and increasing toward the disk. }
\end{figure}

\subsection{Response to black hole growth}
\label{sec:bh}

\noindent
We may also consider the response of the halo to the adiabatic growth
of a central dark object, such as a massive black hole. In this case
the nonlinear problem has already been investigated and solved
numerically by several authors (Peebles 1972; Young 1980; Quinlan et
al. 1995), so the results of our linear calculation are only of
academic interest.

We take the limit of the second of equations (\ref{eq:iso_phi_s}) in which
$r_p\rightarrow 0$, and consider the monopole case $\ell=m=0$.  Only the term
corresponding to the dominant root $\half+x_k=1$ survives since $\Phi^s_{\ell
m}\propto r_p^{x_k-1/2}$ and all other terms have $\Re(x_k)>\half$.  The
total potential is therefore
\begin{equation}
\Phi^t(r)=\left(\Phi^e_{00}+\Phi^s_{00}\right)Y_{00}(\bfOmega)
	=-{G m_p\over r}\left[D'_0(1)\right]^{-1},
\end{equation}
where $D'_0(1)=\half$ so that the linear response amplifies the direct
black-hole potential by a factor of 2.

The interpretation of this result is as follows. At radii $r\lta
r_h\equiv \beta Gm_p$ the black hole induces a (nonlinear) density
cusp in the stellar system. At radii $r\gg r_h$ we might expect a
linear density response, but, as it turns out, the linear terms vanish
so the response is second-order in the black-hole mass (this is
straightforward to show in the case of a system composed of stars on
circular orbits). Thus the total potential perturbation at large radii
consists of the Keplerian potential of the black hole augmented by the
mass in the density cusp. The particular value of the augmented mass
given above, $m_p/D_0'(1)$, is not necessarily accurate since it is
derived from a linear calculation while the cusp is nonlinear.

\section{Discussion}
\label{sec:disc}

\noindent
We have discussed the response of a self-similar stellar system to weak
adiabatic gravitational perturbations. Our problem is idealized because real
stellar systems are only approximately self-similar, and because we neglect
the time-dependence that accompanies most tidal fields, such as those from
orbiting satellites.  Nevertheless, this calculation provides one of the very
few examples where the linear response of an inhomogeneous stellar system can
be explicitly computed in terms of quadratures, and hence offers analytic
insight into the nature of the response. Our results can also be used
to test numerical codes used in linear response calculations.

Our assumptions that the stellar system is self-similar and that the
perturbation is adiabatic can be lifted in more realistic calculations
using matrix methods, such as those of Weinberg (1995, 1997). Weinberg
shows that in some cases individual resonances in the stellar system
can lead to strong enhancements in the external tidal field over a
limited range of radii. 

Using the methods derived above, we have investigated the propagation
of the tidal disturbance from a static satellite into
the inner galaxy; in effect, we have computed the static Love number
for an isothermal stellar system. For odd spherical harmonics (other
than $\ell=1$, for which the adiabatic approximation is not
self-consistent), the response of the stellar system is identical to
the response of the barotropic fluid system with the same density
profile. For even $\ell>0$, we find that the zero-frequency normal modes
of the isothermal sphere, which have the spatial dependence
$\exp(-\alpha\log r)Y_{\ell m}(\bfOmega)$, have complex spatial
eigenvalues $\alpha$ and hence oscillate in $\log r$ at fixed angular
position $\bfOmega$. Partly because of this oscillatory behavior, the
halo response does not strongly amplify the satellite's tidal field in
the inner galaxy: the amplification factor or Love number for the
dominant quadrupole ($\ell=2$) tidal component is never more than
about 2 for $0.01\lta r/r_p\lta 1$, crosses zero near $r/r_p\simeq
0.01$ and is negative over the next decade in radius
(Fig. 2).

Our calculations have some similarity to Young's (1980) investigation of the
response of an isothermal stellar system to the adiabatic growth of a central
black hole. The principal differences are that (i) Young's method is
fully nonlinear, whereas ours is only linear; (ii) our method can be
applied to non-spherical perturbations, such as the slow growth of a disk.

Finally, although we have examined only the singular isothermal stellar system
with density $\rho_0(r)\propto r^{-2}$, it is straightforward to extend these
calculations to other self-similar systems with density $\rho_0(r)\propto
r^{-k}$.

\begin{acknowledgments}
This research was supported by NSERC, the Fund for Astrophysical Research
(C.M.) and an Imasco Fellowship (S.T.).

\end{acknowledgments}

\end{document}